# Damped solitons in an extended nonlinear Schrödinger equation with a spatial stimulated Raman scattering and decreasing dispersion

**E.M. Gromov**[a*] **and B.A. Malomed**[b]

[a]*National Research University Higher School of Economics,*
*25/12 Bolshaja Pecherskaja Ulitsa, Nizhny Novgorod 603155, Russia*
[b]*Department of Physical Electronics, Faculty of Engineering, Tel Aviv University, Tel Aviv, 69978, Israel*

## Abstract

Dynamics of solitons is considered in the framework of an extended nonlinear Schrödinger equation (NLSE), which is derived from a system of the Zakharov's type for the interaction between high- and low-frequency (HF and LF) waves. The resulting NLSE includes a *pseudo-stimulated-Raman-scattering* (pseudo-SRS) term, i.e., a spatial-domain counterpart of the SRS term, which is a known ingredient of the temporal-domain NLSE in optics. Also included is inhomogeneity of the spatial second-order dispersion (SOD) and linear losses of HF waves. It is shown that wavenumber downshift by the pseudo-SRS may be compensated by upshift provided by SOD whose local strength is an exponentially decaying function of the coordinate. An analytical soliton solution with a permanent shape is found in an approximate form, and is verified by comparison with numerical results.

## Highlights

>Dynamics of damped solitons is studied in an extended inhomogeneous nonlinear Schrödinger equation. >We consider media with stimulated-scattering damping acting on low-frequency waves and inhomogeneous spatial dispersion. >An approximate analytical solution for damped solitons is found in an explicit form.

## 1. Introduction

The great interest to the dynamics of solitons is motivated by their ability to travel long distances keeping the shape and transferring the energy and information with no little loss. Soliton solutions are relevant to nonlinear models in various areas of physics which deal with the propagation of intensive wave fields in dispersive media: optical pulses and beams in fibers and spatial waveguides, electromagnetic waves in plasma, surface waves on deep water, etc. [1-7]. Recently, solitons have also drawn a great deal of interest in plasmonics [8-10].

Dynamics of long high-frequency (HF) wave packets is described by the second-order nonlinear dispersive wave theory. The fundamental equation of the theory is the nonlinear Schrödinger equation (NLSE) [11,12], which includes the second-order dispersion (SOD) and cubic nonlinearity (self-phase modulation). Soliton solutions in this case arise as a result of the balance between the dispersive stretch and nonlinear compression of wave packets. In particular, permanent-shape solutions for damped solitons were found in the framework of the NLSE including linear losses of HF waves and spatially-decreasing SOD [4,13].

---
[*] Email address for correspondence: egromov@hse.ru (E.M. Gromov)

The dynamics of narrow HF wave packets is described by the third-order nonlinear dispersive wave theory [1], which takes into account the nonlinear dispersion (self-steeping) [14], stimulated Raman scattering (SRS) [15-17] and third-order dispersion (TOD). The basic equation of the theory is the extended NLSE [17-21]. Soliton solutions in the framework of the extended NLSE with TOD and nonlinear dispersion were found in Refs. [22-29]. In Refs. [30,31], stationary kink waves were found as solutions of the extended NSE with SRS and nonlinear dispersion terms. This solution exists as the equilibrium between the nonlinear dispersion and SRS. For localized nonlinear wave packets (solitons), the SRS gives rise to the downshift of the soliton spectrum [15-17] and eventually to destabilization of the solitons. The use of the balance between the SRS and the slope of the gain for the stabilization of solitons in long telecom links was proposed in Ref. [32]. The compensation of the SRS by emission of linear radiation fields from the soliton's core was considered in Ref. [33]. In addition, the compensation of the SRS in inhomogeneous media was considered in several situations, *viz*., periodic SOD [34,35], shifting zero-dispersion point of SOD [36], and dispersion-decreasing fibers [37].

Intensive short pulses of HF electromagnetic or Langmuir wave in plasmas, as well as HF surface waves in deep stratified water, suffer effective induced damping due to scattering on LF waves, which, in turn, are subject to the action of viscosity. These LF modes are ion-sound waves in the plasma, and internal waves in the stratified fluid. The first model for the damping induced by the interaction with the LF waves was proposed in Ref. [38]. This model gives rise to an extended NLSE with the spatial-domain counterpart of the SRS term, that was call a *pseudo-SRS* one. The equation was derived from the system of the Zakharov's type equations [39,40] for the coupled Langmuir and ion-acoustic waves in plasmas. The pseudo-SRS leads to the self-wavenumber downshift, similar to what is well known in the temporal domain [1,14-17] and, eventually, to destabilization of the solitons. The model elaborated in Ref. [38] also included smooth spatial variation of the SOD, accounted for by a spatially decreasing SOD coefficient, which leads to an increase of the soliton's wavenumber, making it possible to compensate the effect of the pseudo-SRS on the soliton by the spatially inhomogeneous SOD, neglecting the direct effect of the LF-wave loss.

In this work the dynamics of intensive HF wave packets is considered in dispersive nonlinear media, taking into account the scattering on the damped LF waves, linear losses of HF waves, and exponentially decreasing SOD. In the third-order approximation of the nonlinear dispersion-wave theory (for narrow wave packets), the initial Zakharov's-like system of two equations is reduced to an extended inhomogeneous NLSE which includes the pseudo-SRS term, which, as said above, leads to the self-wavenumber downshift. On the other hand, the spatially decreasing SOD causes an increase of the soliton's wavenumber. The equilibrium between the pseudo-SRS and decreasing SOD gives rise to stabilization of the soliton's wavenumber spectrum. Soliton steady-state solutions are found in an explicit approximate form.

## 2. The basic equations and integral relations

We consider the evolution of a slowly varying envelope, $W(\xi,t)$, of the intensive HF wave field in the nonlinear medium with inhomogeneous SOD, taking into account the interaction with the damped LF wave, which is represented by the local perturbation of the effective refractive index, $n(\xi,t)$. The respective system of the Zakharov's type for the unidirectional propagation of the HF and LF waves is [39,40]

$$2i\left(\frac{\partial W}{\partial t}+V\frac{\partial W}{\partial x}\right)+\frac{\partial}{\partial x}\left(q(x)\frac{\partial W}{\partial x}\right)-nW+i\nu W=0, \tag{1}$$

$$\frac{\partial n}{\partial t}+V_S\frac{\partial n}{\partial x}-\delta\frac{\partial^2 n}{\partial x^2}=-\frac{\partial\left(|W|^2\right)}{\partial x}, \tag{2}$$

where $q(x)$ is the SOD, $\nu$ is the linear-losses coefficient of the HF waves, $\delta$ is the viscosity of the LF waves, $V$ is the HF group velocity, and $V_S$ is the velocity of LF waves. As mentioned above, this system may describe intensive Langmuir waves in isotropic plasmas coupled to ion-sound waves, which are subject to the viscous damping. The second-order approximation of the nonlinear dispersion-wave theory corresponds to replacing Eq. (2) by the adiabatic approximation, $n=-|W|^2(V_S-V)^{-1}$, hence envelope $U$ of the HF wave packet obeys the NSLE:

$$2i\frac{\partial W}{\partial t}+\frac{\partial}{\partial \xi}\left[q(\xi+Vt)\frac{\partial W}{\partial \xi}\right]+2\alpha W|W|^2+i\nu W=0,$$

where $\xi\equiv x-Vt$ [accordingly, $q(x)$ is replaced by $q(\xi+Vt)$], and $\alpha\equiv(1/2)(V_S-V)^{-1}$.

In the third-order approximation of the theory (for narrow HF wave packets, with $k\Delta\ll\delta$, where $k$ and $\Delta$ are the characteristic wave number and spatial extension and of the wave packet, respectively), Eq. (2) may be approximated by the nonlinear response of the medium, $n=-|W|^2(V_S-V)^{-1}-\delta(V_S-V)^{-2}\partial\left(|W|^2\right)/\partial\xi$, which leads to the following evolution equation for the HF envelope amplitude:

$$2i\frac{\partial W}{\partial t}+\frac{\partial}{\partial \xi}\left[q(\xi+Vt)\frac{\partial W}{\partial \xi}\right]+2\alpha W|W|^2+\mu W\frac{\partial\left(|W|^2\right)}{\partial \xi}+i\nu W=0, \tag{3}$$

where term $\mu W\partial\left(|W|^2\right)/\partial\xi$, with

$$\mu\equiv\delta(V_S-V)^{-2}, \tag{4}$$

is the spatial counterpart of the SRS effect in the temporal domain. After substitution

$$W\equiv U\exp(-\nu t/2), \tag{5}$$

Eq. (3) takes the form of

$$2i\frac{\partial U}{\partial t}+\frac{\partial}{\partial \xi}\left[q(\xi+Vt)\frac{\partial U}{\partial \xi}\right]+2\alpha U|U|^2\exp(-\nu t)+\mu U\frac{\partial\left(|U|^2\right)}{\partial \xi}\exp(-\nu t)=0. \tag{6}$$

Equation (6) with zero boundary conditions at infinity, $U|_{\xi \to \pm\infty} \to 0$, gives rise to the following integral relations for the field moments:

– the rate of the change of the wave action:

$$\frac{dN}{dt} \equiv \frac{d}{dt} \int_{-\infty}^{+\infty} |U|^2 d\xi = 0;  \quad (7)$$

– the rate of the change of the wave-field momentum:

$$2\frac{d}{dt} \int_{-\infty}^{+\infty} K|U|^2 d\xi = -\mu \exp(-\nu t) \int_{-\infty}^{\infty} \left(\frac{\partial (|U|^2)}{\partial \xi}\right)^2 d\xi - \int_{-\infty}^{\infty} \frac{\partial q}{\partial \xi} \left|\frac{\partial U}{\partial \xi}\right|^2 d\xi; \quad (8)$$

– the rate of the change of the integrated squared gradient of the wave field:

$$\frac{d}{dt} \int_{-\infty}^{+\infty} \left|\frac{\partial U}{\partial \xi}\right|^2 d\xi = -\mu \exp(-\nu t) \int_{-\infty}^{\infty} K\left(\frac{\partial (|U|^2)}{\partial \xi}\right)^2 d\xi$$

$$+ \alpha \exp(-\nu t) \int_{-\infty}^{+\infty} K \frac{\partial (|U|^4)}{\partial \xi} d\xi - \int_{-\infty}^{+\infty} \frac{\partial q}{\partial \xi} \left(\frac{\partial^2 U}{\partial \xi^2} \frac{\partial U^*}{\partial \xi} - \frac{\partial^2 U^*}{\partial \xi^2} \frac{\partial U}{\partial \xi}\right) d\xi; \quad (9)$$

– the rate of the change of the squared gradient of the wave-field intensity:

$$\frac{d}{dt} \int_{-\infty}^{\infty} \left(\frac{\partial (|U|^2)}{\partial \xi}\right)^2 d\xi = 2 \int_{-\infty}^{+\infty} \frac{\partial^2 (|U|^2)}{\partial \xi^2} \frac{\partial (qK|U|^2)}{\partial \xi} d\xi; \quad (10)$$

– the rate of the change of the squared intensity:

$$\frac{d}{dt} \int_{-\infty}^{\infty} |U|^4 d\xi = \int_{-\infty}^{+\infty} qK \frac{\partial (|U|^4)}{\partial \xi} d\xi; \quad (11)$$

– the rate of the change of center-of-mass coordinate:

$$\frac{d}{dt} \int_{-\infty}^{\infty} \xi |U|^2 d\xi = \int_{-\infty}^{+\infty} qK|U|^2 d\xi. \quad (12)$$

In these relations, the complex wave field is represented as $U \equiv |U|\exp(i\phi)$, $U^*$ stands for the complex conjugate, and $K \equiv \partial \phi / \partial \xi$ is the local wavenumber.

## 3. Analytical results
### 3.1. The evolution of field moments

For the analytical consideration of the wave-packet dynamics, we assume that scales of the inhomogeneity of both the SOD term and local wavenumber $K$ are much larger than the spatial width of the wave-packet envelope, $D_{q,K} \gg D_{|U|}$, and approximate the spatial variation of the wavenumber by the linear function, $K(\xi,t) \approx K(\bar{\xi},t) + (\partial K/\partial \xi)_{\bar{\xi}}(\xi - \bar{\xi})$, where $\bar{\xi} \equiv N^{-1} \int_{-\infty}^{+\infty} \xi |U|^2 d\xi$ and $N \equiv \int_{-\infty}^{+\infty} |U|^2 d\xi$. Then we obtain from the imaginary part of Eq. (6), under condition $(\partial |U|/\partial \xi)_{\bar{\xi}} = 0$ (which means that peak of the soliton's amplitude is located at its center):

$$\left(\frac{\partial K}{\partial \xi}\right)_{\bar{\xi}} = -\left(\frac{2}{q|U|}\frac{\partial |U|}{\partial t} + \frac{1}{q}\frac{\partial q}{\partial \xi} K\right)_{\bar{\xi}}. \tag{13}$$

Thus, taking into account Eqs. (6) and (13), for solitary-wave packets the wavenumber distribution can be represented as

$$K(\xi,t) = k(t) + \left(\frac{v}{q} + \frac{V}{q^2}\frac{\partial q}{\partial \xi}\right)_{\bar{\xi}} (\xi - \bar{\xi}), \tag{14}$$

where $k(t) \equiv K(\bar{\xi},t)$. Further, the system of Eqs. (8)-(12) can be cast in the form of evolution equations for parameters of the wave packet:

$$2N\frac{dk}{dt} = -\mu L_0 l \exp(-vt) - q'(\bar{\xi}+Vt)Nz, \tag{15}$$

$$N\frac{dz}{dt} = -\mu L_0 k l \exp(-vt) - 3kq'(\bar{\xi}+Vt)Nz + 2k^3 q'(\bar{\xi}+Vt)N$$
$$- \frac{\alpha}{q(\bar{\xi}+Vt)}M_0\left(v + V\frac{q'(\bar{\xi}+Vt)}{q(\bar{\xi}+Vt)}\right)m\exp(-vt), \tag{16}$$

$$\frac{dm}{dt} = -kq'(\bar{\xi}+Vt)m - \left(v + V\frac{q'(\bar{\xi}+Vt)}{q(\bar{\xi}+Vt)}\right)m, \tag{17}$$

$$\frac{dl}{dt} = -3kq'(\bar{\xi}+Vt)l, \tag{18}$$

$$\frac{d\bar{\xi}}{dt} = kq(\bar{\xi}+Vt), \tag{19}$$

where $q'(\bar{\xi}+Vt) \equiv (\partial q/\partial \xi)_{\bar{\xi}}$, $l \equiv L/L_0$, $m \equiv M/M_0$, $z \equiv Z/N$, $Z \equiv \int_{-\infty}^{\infty} |\partial U/\partial \xi|^2 d\xi$, while $M \equiv \int_{-\infty}^{\infty} |U|^4 d\xi$, $L \equiv \int_{-\infty}^{\infty} (\partial (|U|^2)/\partial \xi)^2 d\xi$, along with the wave action $N$, defined in Eq. (7), are integral characteristics of the wave packet, and $M_0 = M(0)$, $L_0 = L(0)$ are their initial values.

We now select the spatial variation of SOD in the form of $v + Vq'(\bar{\xi}+Vt)/q(\bar{\xi}+Vt) = 0$, corresponding to an exponentially decreasing profile of the SOD,

$$q = q_0 \exp(-vx/V). \tag{20}$$

In particular, the realization of fibers with exponentially decreasing profiles of the SOD was demonstrated experimentally in [41]. Such profiles are created by variation of fiber's diameter. Then system (15)-(19), with the time and the soliton's coordinate redefined as

$$\theta \equiv vt, \eta \equiv v\overline{\xi}/V,  \tag{21}$$

is reduced to

$$2\frac{V}{q_0}\exp\theta\frac{dk}{d\theta} = -pl + z\exp(-\eta), \tag{22}$$

$$\frac{V}{q_0}\exp\theta\frac{dz}{d\theta} = -pkl + 3k\exp(-\eta)z - 2k^3\exp(-\eta), \tag{23}$$

$$\frac{V}{q_0}\exp\theta\frac{dl}{d\theta} = 3kl\exp(-\eta), \tag{24}$$

$$\frac{V}{q_0}\exp\theta\frac{d\eta}{d\theta} = k\exp(-\eta), \tag{25}$$

where $p \equiv \mu V L_0 /(vq_0 N)$. Using first integrals $l = \exp(3\eta)$ and $z = k^2 + (z_0 - k_0^2)\exp(2\eta)$, where $k_0 = k(0)$, $z_0 \equiv Z_0/N$, $Z_0 = Z(0)$, Eqs. (22)-(25) reduce to

$$2\sigma\exp\theta\frac{dy}{d\theta} = -\lambda\exp(3\eta) + y^2\exp(-\eta) + (1 - y_0^2)\exp(\eta), \tag{26}$$

$$\sigma\exp\theta\frac{d\eta}{d\theta} = y\exp(-\eta), \tag{27}$$

where new constants are defined as $\sigma \equiv V/(q_0\sqrt{z_0})$, $y_0 = y(0)$,

$$\lambda \equiv p/z_0 = \mu V L_0/(z_0 v q_0 N), \tag{28}$$

and the rescaled soliton's wavenumber is

$$y \equiv k/\sqrt{z_0}. \tag{29}$$

An equilibrium state of Eqs. (26)-(27) is achieved under conditions

$$k = 0, \ VL_0\mu = q_0 v Z_0. \tag{30}$$

In the equilibrium regime, the wave packet $U$ propagates with the integral moments, $N$, $L_0$, and $Z_0$, keeping their initial values, $N$, $L_0$, $Z_0$, and zero wavenumber. Therefore, the field moments for original wave packet, $W = U\exp(-\theta/2)$ [see Eq. (5)] decay exponentially, $N_W(\theta) = N\exp(-\theta)$, $L_W(\theta) = L_0\exp(-\theta)$, $Z_W(\theta) = Z_0\exp(-\theta)$ (recall that $\theta \equiv vt$). The first integral of these equations is

$$3y^2\exp(-\eta) - \lambda(1 - \exp(3\eta)) + 3(1 - y_0^2)(1 - \exp(\eta)) = 3y_0^2. \tag{31}$$

In Fig. 1, first integral (31) is drawn in the plane of $(y,\eta)$ for $y_0 = 0$ and different values of $\lambda$.

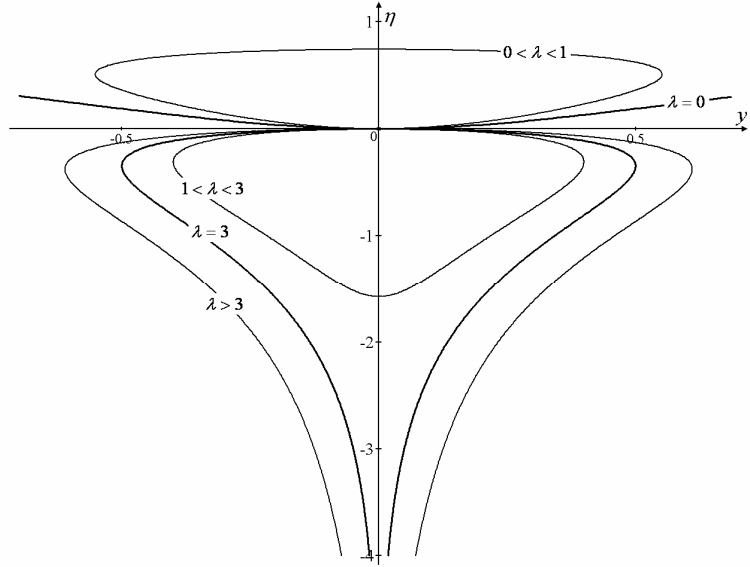

Fig. 1. First integral (31) in the plane $(y,\eta)$ of the soliton's rescaled wavenumber and coordinate [see Eqs. (29) and (20)] for $y_0 = 0$ and different values of constant $\lambda$ [see Eq. (28)]. Point $(0,\ 0)$ - $\lambda = 1$.

The temporal evolution $y(\theta)$ following from Eqs. (26)-(27) is shown in Fig. 2 for initial condition $y_0 = 0$ with different $\sigma$ and $\lambda$.

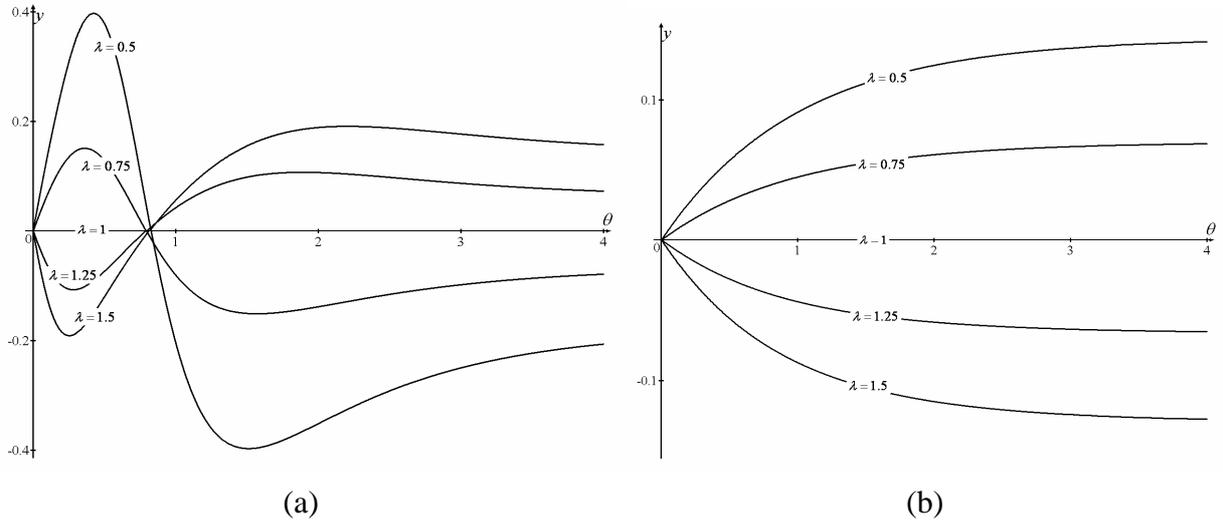

(a)  (b)

Fig.2. Time evolution $y(\theta)$ obtained from Eqs. (26)-(27) for initial condition $y_0 = 0$ with different values of $\sigma$ [(a): $\sigma = 1/10$, (b): $\sigma = 1$], and different $\lambda$.

### 3.2. The soliton solution

We now consider solutions to Eq. (6) for the exponential spatial profile of the SOD defined as per Eq. (20), i.e., with $q(\xi + Vt) = q_0 \exp(-(\xi + Vt)/D)$, in the form of a stationary wave profile, $U(\xi,t) = \psi(\xi)\exp(i\int \Omega(t)dt)$:

$$q_0 \exp\left(-\frac{\xi}{D}\right)\exp\left[t\left(\nu-\frac{V}{D}\right)\right]\frac{d^2\psi}{d\xi^2} - \frac{q_0}{D}\exp\left(-\frac{\xi}{D}\right)\exp\left[t\left(\nu-\frac{V}{D}\right)\right]\frac{d\psi}{d\xi}$$
$$+ 2\alpha\psi^3 - 2\Omega(t)\exp(\nu t)\psi + \mu\psi\frac{d(\psi^2)}{d\xi} = 0. \tag{32}$$

Under conditions $V/D = \nu$ and $\Omega(t) = \Omega_0 \exp(-\nu t)$, Eq. (32) gives

$$q_0 \exp\left(-\frac{\nu\xi}{V}\right)\frac{d^2\psi}{d\xi^2} + 2\alpha\psi^3 - 2\Omega_0\psi - \nu\frac{q_0}{V}\exp\left(-\frac{\nu\xi}{V}\right)\frac{d\psi}{d\xi} + \mu\psi\frac{d(\psi^2)}{d\xi} = 0. \tag{33}$$

Next, we assume that, as said above, the scale of the spatial inhomogeneity of the SOD is much larger for than for the wavepacket's width, $D \equiv V/\nu \gg D_U$. Taking into account that $\varepsilon \sim \nu D_U / V \sim \mu \ll \alpha, q_0$, and expansion $\exp(-\nu\xi/V) \approx 1 - \nu\xi/V$, solution to Eq. (33) is found in the form of

$$\psi = \psi_0 + \psi_1, \text{ with } \psi_1 \sim \varepsilon\psi_0 \ll \psi_0. \tag{34}$$

Keeping terms of order $\varepsilon$, we obtain

$$q_0 \frac{d^2\psi_0}{d\xi^2} + 2\alpha\psi_0^3 - 2\Omega\psi_0 = 0, \tag{35}$$

$$q_0 \frac{d^2\psi_1}{d\xi^2} + \left(6\alpha\psi_0^2 - 2\Omega\right)\psi_1 = \nu\frac{q_0}{V}\frac{d^2\psi_0}{d\xi^2}\xi - \frac{2}{3}\mu\frac{d(\psi_0^3)}{d\xi} + \nu\frac{q_0}{V}\frac{d\psi_0}{d\xi}. \tag{36}$$

Equation (35) gives rise to the fundamental soliton at the zero order,

$$\psi_0 = A_0 \text{sech}(\xi/\Delta), \text{ with } \Delta \equiv \sqrt{q}/(A_0\sqrt{\alpha}) \text{ and } \Omega \equiv \alpha A_0^2/2. \tag{37}$$

Then, after the substitutions of $\eta \equiv \xi/\Delta$ and $\Psi \equiv \psi_1 V/(\nu A_0 \Delta)$, Eq. (36) for the first correction takes the form of

$$\frac{d^2\Psi}{d\eta^2} + \left(\frac{6}{\cosh^2\eta} - 1\right)\Psi = -2\frac{\eta}{\cosh^3\eta} + \frac{\eta}{\cosh\eta} + \frac{5}{4}\frac{\mu}{\mu_*}\frac{\sinh\eta}{\cosh^4\eta} - \frac{\sinh\eta}{\cosh^2\eta}, \tag{38}$$

where the equilibrium value of the strength of the pseudo-SRS term is

$$\mu_* \equiv 5q_0\nu/(8A_0^2 V) \tag{39}$$

[recall that coefficient $\mu$ is defined in Eq. (4)]. In fact, Eq. (39) selects the value of the zero-order soliton's amplitude, $A_0$ [see Eq. (37)], at which the soliton exists for given parameters of the adopted model, i.e., Eqs. (1), (2), and (20). Under condition $\Psi(0) = 0$, Eq. (38) yields an exact solution for the first correction,

$$\Psi(\eta) = \left( \Psi'(0)\tanh\eta - \frac{\eta^2}{4}\tanh\eta + \frac{\mu}{4\mu_*}(\tanh\eta)\ln(\cosh\eta) \right)\mathrm{sech}\,\eta + \frac{1}{12}\left(\frac{\mu}{\mu_*} - 1\right)(\tanh^2\eta)\sinh\eta, \quad (40)$$

cf. a similar solution reported in Ref. [38]. For $\mu = \mu_*$, solution (40) satisfies boundary conditions $\Psi(\eta \to \pm\infty) \to 0$. This solution exists due to the balance between the pseudo-SRS and exponentially decreasing profile SOD. In Fig. 3, distributions $\Psi(\eta)$ for $\mu = \mu_*$ and different values of $\Psi'(0)$ are displayed. At $\mu \neq \mu_*$, solution (40) diverges at the spatial infinity, $|\Psi(\eta \to \pm\infty)| \to \infty$.

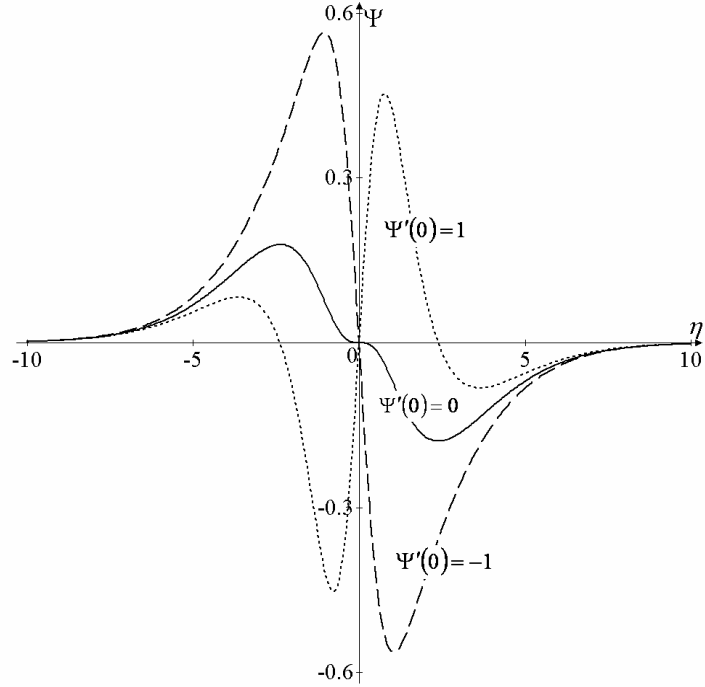

Fig.3. Profiles $\Psi(\eta)$ for $\Psi(0) = 0$, $\mu = \mu_*$, and different values of $\Psi'(0)$.

Note that the full solution, given by Eq. (34), is asymmetric, being a combination of the symmetric and antisymmetric zero- and first-order parts, (37) and (40). Solitons with asymmetric tails arise in the well-known system of linearly coupled NSEs which describes arrays of tunnel-coupled nonlinear optical fibers [42].

## 4. Numerical results

We simulated solutions of the initial-value problem for the wave packet, $U(\xi, t=0) = \mathrm{sech}\,\xi$, in the framework of Eq. (6) for $v/V = 1/10$, $q(\xi) = \exp(-\xi/10)$, $\alpha = 1$ and different values of $V$, $v$ and $\mu$. The analytically predicted equilibrium value of the pseudo-SRS coefficient (39) for this initial pulse is $\mu_* = 1/16$. In direct simulations, the initial pulse for $\mu = 1/16$ is transformed into a stationary localized distribution (the solid curve in Fig. 4) with zero wavenumber. This distribution is close to the analytical solution given by Eqs. (34), (37), and (38), with constants $q_0 = \alpha = A_0 = 1$, $q' = -1/10$, and $\mu = \mu_*$, predicted by Eqs. (35)-(36):

$$|U| = \left(1 + \frac{1}{40}\left((\tanh\xi)\ln(\cosh\xi) - \xi^2 \tanh\xi\right)\right)\mathrm{sech}\,\xi. \qquad (41)$$

In Fig. 4, the profile of the soliton solution (41) is shown by the dotted curve.

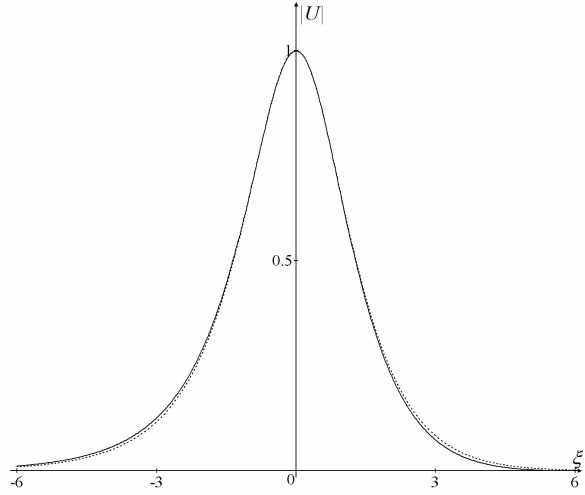

Fig. 4. The solid curve is the result of the numerical solution of Eq. (6) for the soliton's envelope, $|U(\xi)|$, obtained in this stationary form in the time interval $5 < t \leq 400$ for $q(\xi) = \exp(-\xi/10)$ and $\mu = 1/16$. The dotted curve is the analytical solution (41).

At values of the pseudo-SRS coefficient different from $\mu_*$, given by Eq. (39), the simulations produce nonstationary solitons, see an example for $\mu = 2/64 \equiv 0.5\mu_*$ in Figs. 5 and 6.

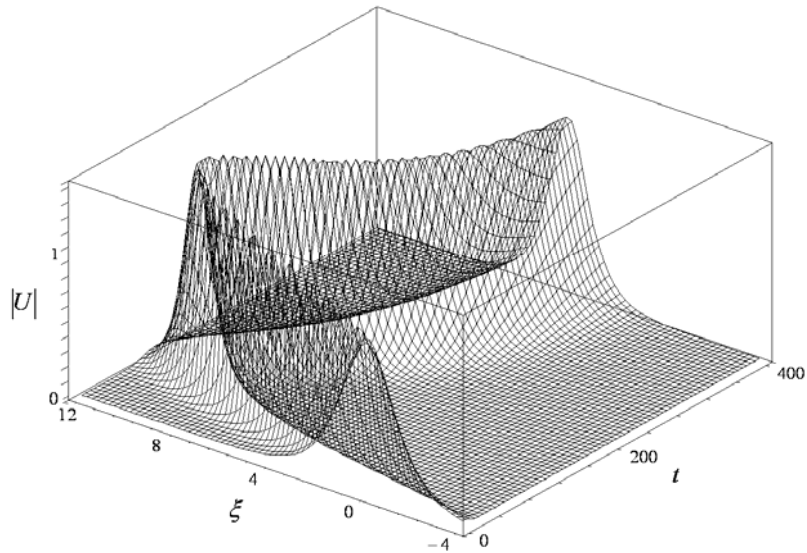

Fig. 5. The numerically simulated evolution of the wave-packet envelope for $\mu = 2/64 \equiv 0.5\mu_*$ and $\nu = 0.01$, $V = 0.1$.

In Fig. 6, numerical results produced, as functions of time, by the simulations for the local wavenumber at the maximum point of the wave-packet's shape, are compared with the analytical counterparts obtained from Eqs. (26)-(27) for different values of $\mu$. Close agreement between the analytical and numerical results is demonstrated by the figure, both for $\mu = \mu_*$, when both the numerically and analytically found wavenumbers remain equal to zero, and for nonstationary pulses at $\mu \neq \mu_*$. A similar picture is observed at other values of the parameters.

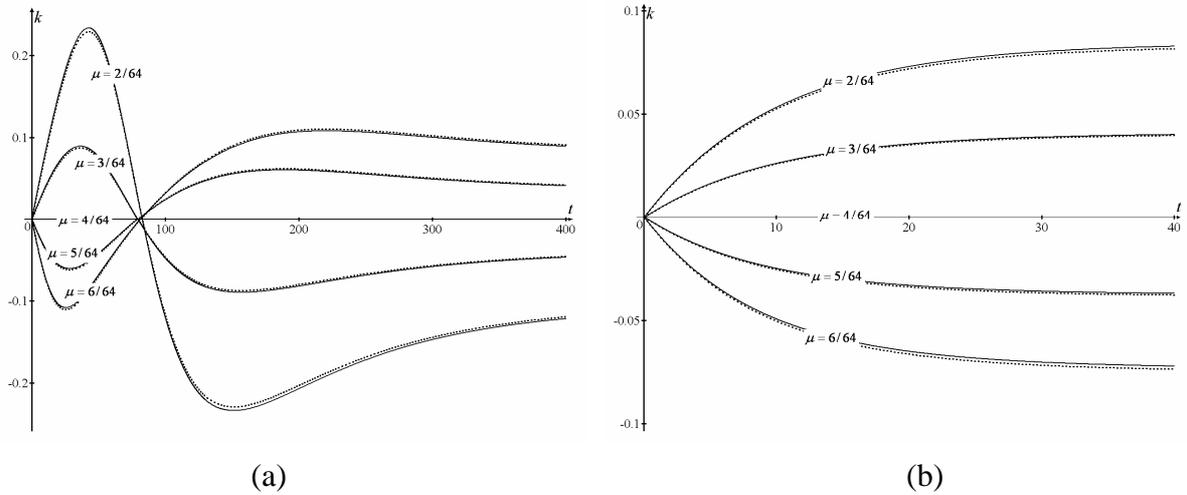

(a)          (b)

Fig.6. Numerical and analytical results (solid and dashed curves) for the local wavenumber at the point of the maximum of the wave-packet envelope versus time for the SOD profile $q(x) = \exp(-x/10)$, see Eq. (20), and different values of $\nu$ and $V$ [(a): $\nu = 0.01, V = 0.1$, (b): $\nu = 0.1, V = 1$], at different values of $\mu$.

## Conclusion

In this work, we studied the soliton dynamics in the framework of the extended inhomogeneous NLSE, derived from the Zakharov's system for the coupled HF and LF (high- and low-frequency) waves. The model includes the pseudo-SRS (pseudo-stimulated-Raman-scattering) effect, the exponentially decreasing SOD (second-order dispersion) and linear losses of the HF waves. The results were obtained by means of analytical method, based on evolution equations for the field moments, and verified by direct simulations. The stationary solitons exist due to the balance between the self-wavenumber downshift, caused by the pseudo-SRS, and the upshift induced by the exponentially decreasing SOD. The analytical solutions are close to their numerically found counterparts.

The present model does not take into regard the nonlinear dispersion and third-order linear dispersion. The compensation of the pseudo-SRS in the model of inhomogeneous media which includes these higher-order terms will be considered elsewhere.

**Acknowledgements**

This study (research grant No 14-01-0023) was supported by The National Research University–Higher School of Economics' Academic Fund Program in 2014/2015.